\documentclass[11pt]{article}
\usepackage{amsmath,epsfig}

\def\kk{ \pmb{k} }
\def\rr{ \pmb{r} }
\def\TT{ \pmb{T} }

\begin{document}

\title{\bf Effective mass of free neutrons in neutron star crust}

\author {\textbf{Nicolas Chamel}\\ \\
Copernicus Astronomical Center (CAMK), Polish Academy of Sciences,  \\
ul. Bartycka 18, 00-716 Warszawa, Poland \\
and \\
LUTH, Paris Observatory, 5 place Jules Janssen, 92195 Meudon, France
}

\maketitle

\bigskip

{\bf Abstract}

The inner layers of a neutron star crust, composed of a Coulomb lattice of neutron rich nuclear clusters immersed in a sea of ``free'' superfluid neutrons,
are closely analogous to periodic condensed matter systems such as electronic, photonic or phononic crystals.
Applying methods from solid state physics to the neutron star context, we study the transport properties of those
 ``free'' neutrons for the outermost layers of the inner crust, near the drip point $\rho_{\rm drip} \sim 4\times 10^{11}$ g.cm$^{-3}$. In particular, we evaluate the
effective neutron mass resulting from Bragg scattering by a band structure calculation. Comparison is made with the case of electrons in solids.
The observational consequences are briefly discussed.

\vskip  1 cm

\section{Introduction}

The outer layers of a neutron star are formed of a solid crust composed of a
Coulomb lattice of very neutron rich nuclei immersed in
a nearly uniform relativistic electron gas. In the deeper layers corresponding to densities
beyond the drip threshold $\rho_{\rm drip}\simeq 4\times 10^{11}\, {\rm g}.{\rm cm}^{-3}$, nuclei are embedded in a sea of ``free''
neutrons which are expected to become superfluids in mature neutron stars whose temperature has dropped below the critical
temperature for the onset of superfluidity (for a review
of neutron star crust matter, see \cite{Hae01} and references therein).
Below the crust, at densities above $\sim 10^{14}\, {\rm g}.{\rm cm}^{-3}$ the nuclei merge into a uniform mixture of nucleons and leptons.

The crust of a neutron star, which represents only about $1\%$ of the mass of the star and $10\%$ of its radius,
is nevertheless important to understand observational phenomena, such as for instance pulsar glitches. Pulsars are
strongly magnetised rotating neutron stars whose period, ranging from milliseconds to a few seconds, very
slowly increases due to the loss of energy by electromagnetic and gravitational radiation. However some pulsars have been
observed to suddenly spin up. These so called ``glitches'' are characterised by a frequency jump, which varies from
$\delta \Omega/\Omega \sim 10^{-9}$ (Crab) up to
$\delta \Omega/\Omega \sim 10^{-6}$ (Vela), often accompanied by a permanent change in the slow down rate
from $|\delta \dot \Omega/\dot \Omega| \sim 10^{-5}-10^{-4}$ (Crab) to  $10^{-3}-10^{-2}$ (Vela). The relaxation following a glitch lasts
from days to years. It was suggested by Baym \textit{et al.} \cite{Baym69} soon after the discovery of the first pulsars that
such very long timescales indicate the presence of superfluid components in the interior of the star.
It is now widely accepted that pulsar glitches are due to a sudden tranfer of angular momentum from the faster rotating neutron superfluid
to the solid crust. However the detailed mechanism of these glitches remain uncertain. In order to understand such phenomena, it is
necessary to investigate the transport properties of the neutron superfluid in the inner crust \cite{CCHII, ChamelCarter05}.

It has been recently shown that the effects of the
nuclear clusters on the superfluid neutrons lead to a renormalisation
of the neutron mass due to Bragg scattering \cite{CCHI}, which gives rise to the entrainment effect
according to which the momentum of the neutron superfluid is not aligned with the corresponding velocity \cite{CCHII}.
This effective mass has been calculated in the very deep layers of the crust
close to the interface with the liquid core for spherical nuclei \cite{Chamel05} and for the ``pasta'' phases
where the crust is formed by a lattice of slab (``lasagna'') or cylinder (``spaguetti'') shaped
nuclei \cite{CCHI}. The effective mass has been found to be close to the bare neutron mass in the bottom of the crust while
at lower densities the effective mass is strongly increased, reaching values as high as $m_\star \sim 15 m_n$ at
the baryon density $5 \times 10^{13}$ g.cm$^{-3}$. The purpose of the present work is to extend this
calculation to much lower densities near the neutron drip point.

\section{Band theory and effective mass}

Since the classical paper of Negele \& Vautherin \cite{NV73} (subsequently refered simply as N\&V), quantum calculations
in the inner crust have been carried out within the Wigner-Seitz (W-S) approximation according to which the crust is
decomposed into a compact arrangement of independent spheres centered around each nuclear cluster. While this approach
greatly simplifies the determination of the equilibrium structure of the crust and its static properties such as the equation of state
for instance, it cannot be applied for the study of transport properties. Since the free neutrons in the inner crust of a neutron star
are closely analogous to free electrons in metals, a natural step forward beyong the W-S approximation
is to apply the well know band theory.

This theory is not only restricted to the description of electrons in solids, but has been also applied to photonic and phononic crystals
in which light (photons) and sound (phonons) respectively play the role of the electrons.
Within the band theory, the crust is assumed to be a perfect lattice whose composition is known. The clusters forming the crust
are treated classically (this approximation is justified since each cluster contains about one hundred nucleons or more) while
the free neutrons are considered as independent quasiparticles in
some effective mean field which takes into account the interactions with the other nucleons (in the superfluid phase
or in the clusters).

\subsection{Effective neutron mass}

Subject to a more or less arbitrary convention specifying the density $n_f$ of ``free'' neutrons, the effective neutron mass can then
be expressed as \cite{CCHI}
\begin{equation}\label{effmass}
m_\star = \frac{n_f}{\cal K} \, .
\end{equation}
Within the BCS approximation, the ``mobility'' coefficient $\cal K$  is given by an integral over
the first Brillouin zone (B-Z)  \cite{CCHIII}
\begin{equation}\label{BCSmobility}
{\cal K} = \frac{1}{12\pi^3 \hbar^2} \sum_\alpha  \int_{\rm BZ}  |\boldsymbol{\nabla}_{\kk} {\cal E}_{\kk\alpha}|^2
\frac{\Delta_F}{\sqrt{({\cal E}_{\kk\alpha}-\mu)^2 + \Delta_F^2}} {\rm d}^3 \kk\, ,
\end{equation}
where ${\cal E}_{\kk\alpha}$ is the single particle energy of Bloch wave vector $\kk$ and band index $\alpha$,
$\Delta_F$ is the Fermi surface averaged value of the pairing gap and $\mu$ is the Fermi energy.

At the densities of the inner crust region near the drip point for which we are interested, the pairing gaps obtained in microscopic
calculations of homogeneous pure neutron matter are very small compared to the typical Fermi energies \cite{Baldo05} and since the mobility
coefficient is determined mainly by the single particle
energies ${\cal E}_{\kk\alpha}$, we shall neglect the pairing for the calculation of the effective neutron mass.
In this case the equation reduces
to a Fermi surface integral \cite{CCHI}
\begin{equation}\label{mobility}
{\cal K} = \frac{1}{12\pi^3 \hbar^2} \sum_\alpha \int_{\rm F}  | \boldsymbol{\nabla}_{\kk} {\cal E}_{\kk\alpha} | {\rm d} S^{(\alpha)}\, .
\end{equation}

The effective mass $m_\star$, as defined by (\ref{effmass}) and (\ref{mobility}),
has been refered in the solid state physics literature as an ``optical'' effective (electron) mass since it appears
in the calculation of the real part of the dielectric constant of cubic metals \cite{Cohen58}, which can be written in the regime
$\omega \tau \gg 1$($\tau$ being the relaxation time) as
\begin{equation}
\varepsilon\{ \omega\} \simeq 1 - \frac{\omega_{p \star}^2}{\omega^2} +\varepsilon_{\rm inter}
\end{equation}
where $\omega_{p \star}$ is an effective plasma frequency defined in terms of the density $n_f$ of ``free'' electrons and of their
effective mass $m_\star$ by
\begin{equation}
\omega_{p \star}^2 = \frac{4\pi e^2 n_f}{m_\star}
\end{equation}
and $\varepsilon_{\rm inter}$ is the contribution due to interband transitions which is essentially constant for sufficiently low frequencies.
It can be seen from (\ref{effmass}) that this plasma frequency can be expressed as $\omega_{p \star}^2 =4\pi e^2 \cal K$ and
 is therefore independent of the arbitrary convention used to specify which electrons are ``free''.

It should also be remarked that the electric conductivity $\sigma$ is related to the mobility
coefficient $\cal K$ introduced previously by the simple formula
\begin{equation}\label{conductivity}
\sigma= e^2 \tau \cal K \, ,
\end{equation}
where $e$ is the electron charge and $\tau$ is the relaxation time. 
Unlike the effective electron mass $m_\star$,
the electric conductivity (\ref{conductivity}) is obviously independent of the convention adopted for defining
which electrons are considered to be ``free''. This point has been usually somehow obscured in the litterature
by writing the conductivity in the form $\sigma=n_f e^2 \tau/m_\star$ of the classical
Drude's model.

This effective mass can therefore be deduced experimentally by studying the optical properties of solids or measuring the electric 
conductivity (or equivalently the resistivity $\rho=1/\sigma$) provided the relaxation time $\tau$ is known. 
Values are typically in the range $m_\star \sim 1-2 m_e$
for most elements except for beryllium which has a very small effective mass $m_\star/m_e=0.46$ and
lead whose estimates of effective mass reach $m_\star/m_e=2.71$ as reported in reference \cite{Huttner96} (see appendix
for some numerical values of $m_\star$). 
The density $n_f$ of ``free'' electrons is defined by counting for each atom in the crystal the number of electrons 
in the outermost atomic shell. For instance monovalent metals (such as alkali or noble metals) have only one ``conduction'' or ``free'' 
electron per atom in the crystal therefore $n_f=1/{\cal V}_{\rm cell}$ where ${\cal V}_{\rm cell}$ is the volume of the W-S cell.

The mobility coefficient (\ref{mobility}) can also be written as the integral over the Fermi volume
\begin{equation}
{\cal K} = \frac{1}{3} \sum_\alpha \int_{\rm F} {\rm Tr} \left\{ \frac{1}{m_\star \{ \kk \}} \right\} \, ,
\end{equation}
of the trace of the \emph{local} effective mass tensor \cite{Kittel} defined by
\begin{equation}\label{effmasstensor}
 \frac{1}{m_\star \{ \kk \}}_{ij} \equiv \frac{1}{\hbar^2}\frac{\partial^2 {\cal E}_{\kk}}{\partial k_i \partial k_j} \, .
 \end{equation}
 While the components of this tensor can be either positive or negative, it is clear from (\ref{BCSmobility}) and (\ref{mobility}) that the
 ``optical'' effective mass defined by (\ref{effmass}) is always positive.

Let us mention that the concept of the \emph{local} effective mass tensor, which is well known in solid state physics,
has been introduced in the context of neutron diffraction rather recently \cite{Zeilinger86}. The effective mass
for neutrons propagating along Bragg planes is found to be very small compared to the bare mass by several orders of
magnitude as expected from the analogy with electrons in solids (the electron effective mass at the edge of
the band gap in semiconductors is reduced by about two orders of magnitude \cite{Kittel}).
On the other hand,  the ``optical'' effective mass $m_\star$ as defined by (\ref{effmass}) is an average of the local
effective mass (\ref{effmasstensor}) over all occupied states, not only those for which
the Bloch wave vector lies on a Bragg plane, and can therefore be expected to be much larger (it has already been pointed out
that the electron effective mass is indeed usually increased in metals, see also appendix).

\subsection{Band model of free neutrons}
The structure of the inner crust has been determined by N\&V
within a density matrix expansion method whose equations are similar to those obtained with effective two body nucleon-nucleon interactions
of the Skyrme type in the Hartree-Fock
approximation. We take their results for the lattice spacing and the composition
of the nuclear clusters. The equilibrium structure of the crust is formed of a body centered cubic lattice.

From the self-consistent calculations of N\&V, we construct an effective single particle
equation for the free neutrons, neglecting spin-orbit coupling as was done by N\&V (this approximation can be justified
from the fact that the nuclear clusters, which are very neutron
rich, have a very diffuse surface)

\begin{equation} \label{bandmodel}
\hat H \varphi_{\kk\alpha}   = {\cal E}_{\kk\alpha} \varphi_{\kk\alpha}  \, ,
\end{equation}

\begin{equation}\label{Hamiltonian}
 \hat H = -\boldsymbol{\nabla} \cdot \frac{\hbar^2}{2 m_n^{_\oplus}\{ \rr \}} \boldsymbol{\nabla} + U_n\{ \rr \} \, .
\end{equation}

As a result of the lattice symmetry, the single particle equations have to satisfy the Floquet-Bloch theorem \cite{Kittel}
\begin{equation} \label{Bloch}
\varphi_{\kk\alpha}  \{ \rr +\TT \} = e^{{\rm i}\, \kk\cdot \TT} \varphi_{\kk\alpha} \{ \rr  \}
\end{equation} where $\TT$ is any lattice translation vector and $\bf k$ is a wave vector belonging
to the first BZ. The equation can therefore be solved inside the W-S cell of the lattice (a truncated octahedron)
with the boundary conditions (\ref{Bloch}). In the W-S approximation, this cell is replaced by a sphere with more or less arbitrary boundary
conditions. For instance, N\&V imposed the vanishing of the wave function or its radial derivative on the sphere depending on its parity.
In this approximation, the $\kk$ dependence of the single particle states (resulting from the translational symmetry) is therefore neglected.

The kinetic term $-\hbar^2/2m_n^{_\oplus}\{ \rr \}$ and the mean field potential $U_n\{ \rr \}$ are deduced from
the self-consistent neutron and proton densities $n_n\{ \rr \}$ and $n_p\{ \rr \}$ obtained by N\&V,
by applying the extended Thomas-Fermi approximation up to second order
with Skyrme interactions \cite{Brack85}

\begin{equation}
\frac{\hbar^2}{2m_n^{_\oplus}\{ \rr \}} = \frac{\hbar^2}{2m_n} + \frac{1}{8}\biggl(t_1(2+x_1)+t_2(2+x_2)\biggr) n_b\{ \rr \} - \frac{1}{8}\biggl(t_1(1+2 x_1)-t_2(1+2x_2)\biggr) n_n\{ \rr \} \, ,
\end{equation}

\begin{eqnarray}
 \lefteqn{ U_n\{ \rr\}  = \frac{1}{2} t_0 \bigl[ (2+x_0) n_b\{ \rr\} -(1+2 x_0) n_n\{ \rr\} \bigr]-\frac{1}{24} t_3 \bigl[ (2+x_3)(2+\gamma) n_b\{ \rr \}^{\gamma+1} } \nonumber\\
  & - & (2 x_3+1)(2 n_b\{ \rr\}^\gamma n_n\{ \rr\}+\gamma n_b\{ \rr\}^{\gamma-1} (n_p\{ \rr\}^2+n_n\{ \rr\}^2))\bigr] \nonumber\\
  &+& \frac{1}{8} \bigl[ t_1(2+x_1)+t_2(2+x_2)\bigr] \tau\{ \rr\}+\frac{1}{8}\bigl[ t_2(2x_2+1)-t_1(2x_1+1)\bigr] \tau_n\{ \rr\} \nonumber  \\
  &+&  \frac{1}{16} \bigl[ t_2 (2+x_2)-3 t_1(2+x_1)\bigr] \nabla^2 n_b\{ \rr\} \nonumber \\
  &+ & \frac{1}{16}\bigl[ 3 t_1(2 x_1+1)+t_2 (2 x_2+1)\bigr] \nabla^2 n_n\{ \rr\}
\end{eqnarray}
where the kinetic densities are given by ($q=n,p$ for neutron and proton respectively)
\begin{equation} \tau_q = \tau_q^{_{\rm TF}} + \tau_q^{(2)} \, ,
\end{equation}
where $\tau_q^{_{\rm TF}}$ is the well known Thomas-Fermi expression
\begin{equation} \tau_q^{_{\rm TF}} = \frac{3}{5} (3\pi^2)^{2/3} n_q^{5/3}
\end{equation}
and the second order contribution $\tau_q^{(2)} = \tau_q^{_{\rm BW}}+ \tau_q^{_{\rm NL}}$
is given by the Bethe-Weisz\"acker term
\begin{equation} \tau_q^{_{\rm BW}} =\frac{1}{36} \frac{(\boldsymbol{\nabla} n_q)^2}{n_q} + \frac{1}{3}\Delta n_q \, ,
\end{equation}
plus non local terms arising from the Skyrme masses $f\equiv m/m_q^\oplus$ (calculated here for pure matter of the species $q$)
\begin{equation}  \tau_q^{_{\rm NL}} =  \frac{1}{6} \frac{\boldsymbol{\nabla} n_q \cdot \boldsymbol{\nabla} f}{f} + \frac{1}{6} n_q \frac{\Delta f}{f}
- \frac{1}{12} n_q \biggl(\frac{\boldsymbol{\nabla} f}{f}\biggr)^2 \, .
\end{equation}

We have used the SkM \cite{SkM} parametrisation for the Skyrme force which is expected to give
results close to those of N\&V since its parameters like those in the density matrix expansion method of N\&V,
were adjusted on the neutron matter calculation of Siemens \& Pandharipande \cite{Siemens71}
and since this force was used a few years later by Bonche \& Vautherin \cite{Bonche81} in the case of hot neutron star matter.
For comparison, we have also used the more recent
parametrisation SLy4 \cite{SLy4, SLy4b, SLy4c}, which has been specifically constructed for astrophysical purposes and which has been already applied in neutron
star crust calculations \cite{Douchin00, DouchinHaensel00, Magier2003, Sandulescu04, Sandulescu04b, Khan05}.
The parameters of these two Skyrme interactions are reported in the appendix.

\section{Numerical methods}

\subsection{LAPW method}

We have solved the equation (\ref{bandmodel}) by the standard solid state physics LAPW method \cite{Singh94}.
This method, whose idea was first introduced a long time ago by Slater \cite{Slater37}
for the treatment of electrons in solids, provides a natural extension of the W-S approximation by combining
the spherical symmetry aroung each nuclear cluster with the translational symmetry of the lattice.

The W-S cell is decomposed into two regions: a sphere of radius $R_s$
surrounding the cluster and an intersticial region, where in the muffin tin approximation
the Skyrme mass and the mean field are assumed to take uniform values.

The equation (\ref{bandmodel})
is solved by expanding the wave function on a mixed basis set
\begin{equation}\label{basis}
\varphi_{\kk}\{ \rr \} = \sum_\alpha c_\alpha \phi_{\alpha \kk} \{ \rr \} \, .
\end{equation}
In the intersticial region the basis functions are defined by plane waves
\begin{equation}
\phi_{\alpha \kk}\{ \rr \} = \frac{1}{\sqrt{{\cal V}_{\rm cell}}} e^{ {\rm i}\, (\kk+\pmb{K}_\alpha)\cdot \rr} \, ,
\end{equation}
where $\pmb{K}_\alpha$ are reciprocal lattice vectors,
which ensures that the Floquet-Bloch theorem (\ref{Bloch}) is satisfied. Inside the Slater sphere, the basis functions are given by
\begin{equation}\label{eq.3}
\phi_{\alpha\kk}\{ \rr \} = \sum_{\ell=0}^{\ell_{\rm max}} \sum_{m=-\ell}^\ell \biggl( A_{\ell m } u_\ell\{ r \} + B_{\ell m} \dot{u}_\ell\{r\} \biggr)Y_{\ell m}\{ \hat r\}
\end{equation}
where $Y_{\ell m}$ are spherical harmonics ($\hat r$ refers to the angular coordinates of $\rr$) and
$u_\ell$ is the radial solution of the equation (\ref{bandmodel}) inside the sphere for \emph{a given energy} ${\cal E}_\ell$.
The dot means derivative with respect to the energy. The coefficients $A_{\ell m}$ and $B_{\ell m}$ are determined by
imposing the continuity of the wave function and of its gradient on the sphere (see \cite{Singh94} for
further details).

From the Rayleigh-Ritz variational theorem, it follows that the Schr\"odinger equation
(\ref{bandmodel}) reduces to a generalised eigenvalue problem
\begin{equation}
\sum_\beta H_{\alpha\beta} c_\beta = {\cal E} \sum_\beta S_{\alpha\beta} c_\beta \, ,
\end{equation}
where $H_{\alpha \beta}$ is the matrix of the Hamiltonian (\ref{Hamiltonian})
\begin{equation}
H_{\alpha\beta} = \langle \phi_{\alpha \kk} | \hat H \phi_{\beta \kk} \rangle \, ,
\end{equation}
and $S_{\alpha\beta}$ is the overlap matrix
\begin{equation}
S_{\alpha\beta} = \langle \phi_{\alpha \kk} | \phi_{\beta \kk} \rangle \, .
\end{equation}


The expansion (\ref{basis}) is truncated such as to include all reciprocal lattice vectors satisfying
\begin{equation}\label{cutoff}
\frac{\hbar^2 (\kk+{\pmb{K}_\alpha})^2}{2 m_n} < E_{\rm cutoff} \, .
\end{equation}
The choice of the radius $R_s$ of the Slater sphere and of the linearisation energies $\{{\cal E}_\ell\}$ is rather delicate
due to the occurence of ``ghost'' bands. These bands can be however easily identified from
their low $\kk$ dependence (``flat'' bands in the energy spectrum) and from their high sensitivity to small changes
of the parameters $R_s$ and ${\cal E}_\ell$ (for a discussion on this issue, see for instance \cite{Singh94}).

\subsection{Fermi surface integration}

Fermi surface integrations were carried out within the Gilat-Raubenheimer scheme \cite{GilatRaubenheimer66}. The first B-Z
is decomposed into small identical cubes, inside which the Fermi surface (if any) is approximated by a plane.
From the knowledge of the energy bands and their gradient, it can be shown that the integral can be performed analytically inside each of those cubes.
The integral over the whole Fermi surface is then obtained by summing the contributions of all the cubes. We have computed
the energy gradient $\boldsymbol{\nabla}_{\kk} {\cal E}_{\kk}$ by a finite difference scheme. It seems at first sight that
the application of the Hellmann-Feynman theorem \cite{Feynman39}, from which the gradient is obtainable by the formula
\begin{equation}\label{gradientE}
\boldsymbol{\nabla}_{\kk} {\cal E}_{\kk} = \sum_{\alpha, \beta} c^*_\alpha c_\beta \biggl(\boldsymbol{\nabla}_{\kk} H_{\alpha \beta}
 - {\cal E}_{\kk} \boldsymbol{\nabla}_{\kk} S_{\alpha \beta} \biggr) \, ,
\end{equation}
would lead to more accurate results.
However this method, which we already applied in previous works \cite{CCHI, Chamel05},
yield poor results in the present case because the errors in the wave functions obtained from the LAPW method are of order of $({\cal E}-{\cal E}_\ell)^2$,
while the errors in the energy bands are only of order $({\cal E}-{\cal E}_\ell)^4$ \cite{Koelling75}.

We have tested the code by computing
the energy bands and the effective mass $m_\star$ in the empty lattice model \cite{Shockley37}.
For comparison, we have also computed the energy bands and ``optical'' effective electron mass of copper,
which has been very often used as a benchmark for band structure calculation methods.

\subsection{Application to copper}
\label{Copper}

\begin{figure}
\centering
\epsfig{figure=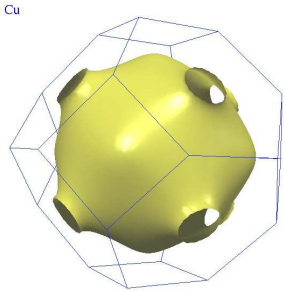, height=4 cm}
\caption{Fermi surface of copper obtained by Choy \textit{et al.} \cite{Choy00} (available online at \textit{http://www.phys.ufl.edu/fermisurface/}).}
\label{fig.Cu}
\end{figure}

Copper is a noble metal whose electronic atomic configuration is $[Ar] 3d^{10}4s^1$ ($[Ar]$ denotes the electronic configuration of argon).
A typical feature of such crystals is the mixing or ``hybridization'' of $s$ and $d$ atomic shells which
leads to significant distortions of the Fermi surface from a sphere as shown on figure \ref{fig.Cu}
(in contrast to alkali metals for which the Fermi surface is nearly spherical).

The closed atomic shells $1s$, $2s$, $2p$, $3s$ and $3p$ give rise
to fully occupied dispersionless energy bands in the crystal (``core'' states) and it is therefore necessary to consider
only the bands arising from the ``valence'' electrons in the $3d$ and
$4s$ atomic shells. We have computed the energy bands of copper employing
 the classical Chodorov potential (see for instance
\cite{Burdick63} and references therein). Copper has a face centered cubic structure with a lattice spacing $a=3.6147$ \AA.

\begin{figure}
\centering
\epsfig{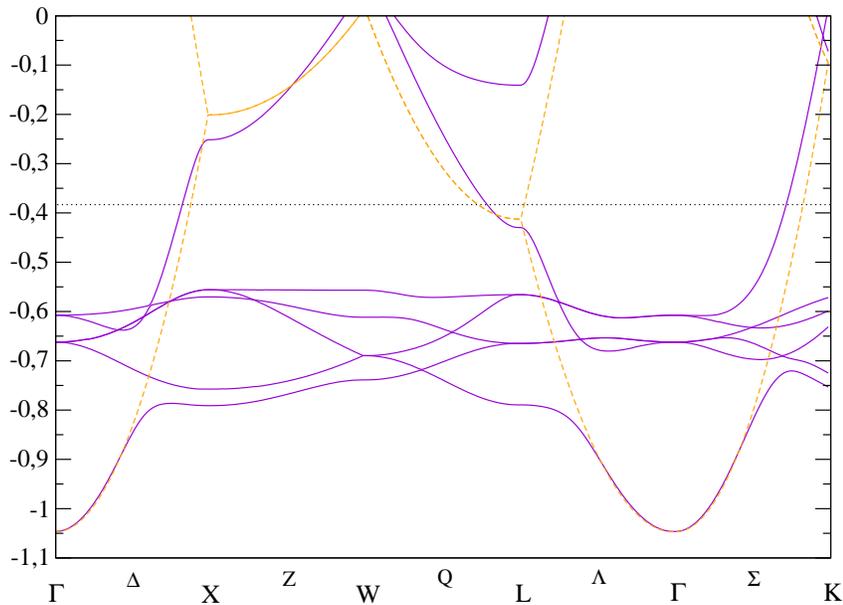}
\caption{Band structure of copper (energies are shown in Rydbergs) along high symmetry lines of the first B-Z (with the conventional labelling \cite{Koster57}),
computed with the LAPW method (see the text for further details). For comparison we have also plotted the energy spectrum (dashed line) of the ideal Fermi gas
suitably translated into the first B-Z. The horizontal dotted line indicates the position of the Fermi energy.}
\label{fig.Cu_bands}
\end{figure}

While the muffin-tin approximation is very good in neutron star crust, it is not so good in ordinary solids due to the long range
of the Coulomb force. The radius $R_s$ of the Slater sphere is thus set at the maximum possible value, namely half the nearest
neighbor distance $R_s=\sqrt{2}a/4$. As already discussed by Koelling \& Arbman \cite{Koelling75}, the bands are rather
sensitive to the choice of the linearisation energies ${\cal E}_\ell$ due to the presence of narrow (``localized'')
$d$ bands with low $\kk$ dispersion. We thus calibrated the energies ${\cal E}_\ell$ by computing the band structure
with the original non linear APW method \cite{Slater37}. This method is more accurate but is computationally much more
expensive than the linearised version. The expansion of the basis functions was truncated to $E_{\rm cutoff}=15$ 
Ryds and $\ell_{\rm max}=4$. We found a fairly good agreement between the two methods with the following setting (energies are in Ryd):
${\cal E}_0=-0.90$, ${\cal E}_1=-0.20$, and ${\cal E}_2={\cal E}_3={\cal E}_4=-0.70$. As can be seen on figure \ref{fig.Cu_bands}
the energy bands show some resemblance to the ideal Fermi spectrum except for  low $\kk$ dispersion $d$ bands in the range 
$-0.55 < {\cal E} < -0.8$ Ryds. This is a characteristic feature of transition elements. 

We have used the Chodorov potential to compute the mobility coefficient of copper and the associated electron (``optical'') effective mass according to
the equations (\ref{effmass}) and (\ref{mobility}). 
For this calculation we have fixed all the linearisation energies to the Fermi energy ${\cal E}_\ell = \mu=-0.385$ Rydbergs.
In solid state physics, the number of
``free'' electrons is usually taken as the number of electrons in the ``conduction'' band. In the case of copper, there is one ``conduction''
electron per ion in the crystal from the atomic $4s$ shell and therefore $n_f=1/{\cal V}_{\rm cell}$. 
The corresponding (``optical'') effective electron mass as defined by (\ref{effmass}) and (\ref{mobility}),
$m_\star \simeq 1.285 m_e$ obtained by integration over 11375 cubes in the irreducible domain (therefore 546000 cubes in the B-Z),
is in agreement with the earlier estimate of Seagall \cite{Seagall61} for the same potential $m_\star/m_e=1.3 \pm 0.1$.
There is however some ambiguities in the specification of the ``free'' electron density $n_f$ in the case of transition metals since these
elements may have different possible valences due to the presence of partially or completely filled $d$ shell in the atom. 
We have also seen in the case of copper(see figure \ref{fig.Cu_bands}) that in the crystal the electrons 
from the $3d$ atomic shell strongly hybridize with the ``conduction'' electrons
from the outermost $4s$ shell. One may therefore adopt the convention of counting also these $d$ electrons as ``free'',
in which case the effective mass becomes much larger $m_\star \simeq 14.14 m_e$ since $n_f=11/{\cal V}_{\rm cell}$.

\section{Results and discussion}

We have considered the largest W-S cell of N\&V (the radius
of the W-S sphere is $R_{_{\rm WS}}= 54$ fm) which corresponds
to the onset of neutron drip at the baryon density
$\rho=4.66 \times 10^{11}$ g.cm$^{-3}$.
The nucleon densities are expressed as Woods-Saxon functions
inside the Slater sphere of radius $R_S$
\begin{equation}
n_q\{ r \} = n_{q o} + (n_{q i}-n_{q o})/\biggl(1+e^{(r-R_q)/\xi_q}\biggr) \, ,
\end{equation}
and are constants outside (and equal to $n_q\{ R_s \}$).
The parameters were adjusted so as to fit the density profiles given by N\&V, imposing that
$n_{po}=0$ and $n_{no}=\tilde\rho_G$ where $\tilde\rho_G$ is the average density
 of the exterior neutron gas given by N\&V.
The lattice spacing $a$ of the body centered cubic lattice was defined such that the volume of the W-S polyhedron, which
is given by ${\cal V}_{\rm cell}=a^3/2$ is equal to the volume of the W-S sphere, which leads to the formula
\begin{equation}
a=2 (\pi/3)^{1/3} R_{_{\rm WS}} \, .
\end{equation}

We found good convergence with the energy cutoff ${\cal E}_{\rm cutoff}=1$ MeV and $\ell_{\rm max}=6$. We
fixed the radius of the Slater sphere to $R_s=35$ fm and all energy parameters  ${\cal E}_\ell$ for $0\leq \ell \leq \ell_{\rm max}$
 to the mid value $0.15$ MeV for the range $0<{\cal E}<0.3$ MeV (the energy origin is chosen as the constant
 value of the potential outside the muffin tin). We compared the band structure
 with that of the non linear APW method \cite{Slater37} and we found good agreement with this choice of parameters.

\begin{figure}
\centering
\epsfig{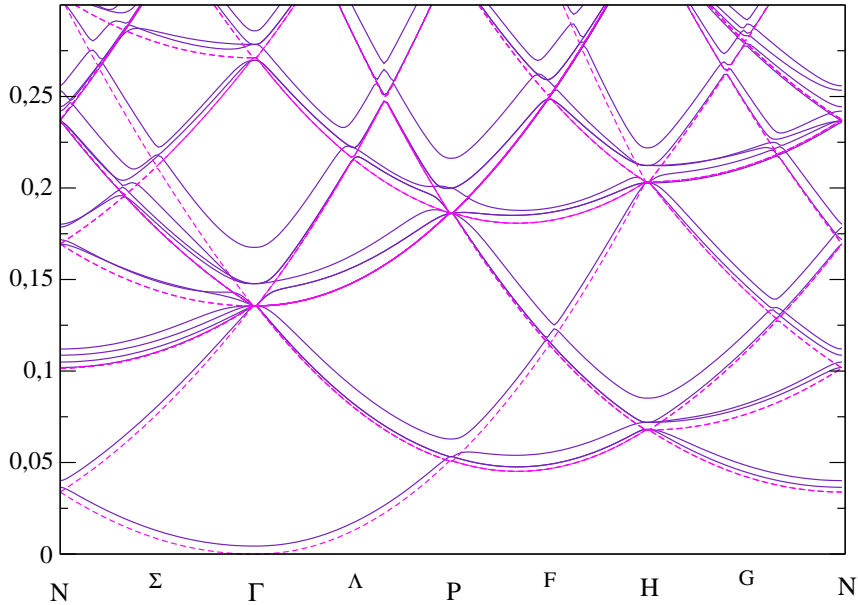}
\caption{Energy bands (in MeV) of the ``free'' neutrons in neutron star crust (solid line) along high symmetry lines of the first B-Z (using the
conventional labelling \cite{Koster57})
compared to the energy spectrum of the ideal Fermi gas (dashed line) suitably translated into the first B-Z for comparison. The band structure has been
obtained with SkM Skyrme forces (details are explained in the text).}
\label{fig.bands}
\end{figure}

\begin{figure}
\centering
\epsfig{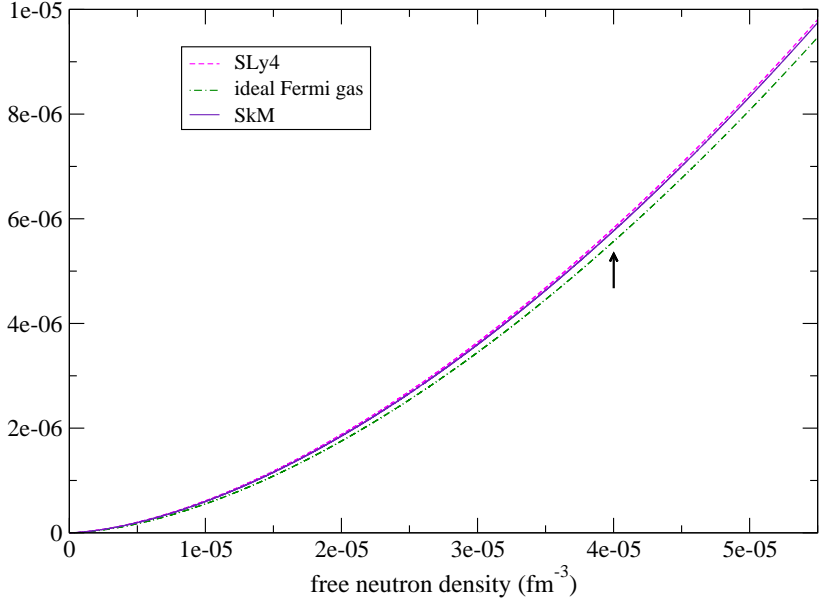}
\caption{Total energy density $\varepsilon$ (in MeV per fm$^{3}$) of the free neutrons as a function of the free neutron density $n_f$ (in fm$^{-3}$)
compared with that of the ideal Fermi gas. The arrow indicates the actual free neutron density in the W-S cell considered by Negele \& Vautherin.}
\bigskip
\label{fig.eos}
\end{figure}

The single particle energy spectrum for the ``free'' neutrons is shown on figure \ref{fig.bands} for wave vectors $\kk$ along specific symmetry lines
of the first B-Z. The figure shows that unlike the band structure of copper computed in the previous section (see figure \ref{fig.Cu_bands},
the neutron energy spectrum is actually very close to that of an ideal Fermi gas, except for avoided crossings around some specific high symmetry points.
The reason of this similarity lies
in the so called Phillips-Kleinman cancellation theorem \cite{PhillipsKleinman59}. As a
result of the Pauli exclusion principle, the orthogonalization of the ``free'' neutrons states to the ``core'' states leads to a repulsive
non local and energy dependent potential which essentially vanishes outside nuclei and which partially compensates the strong attractive
nuclear potential inside nuclei. In the present case, as the nuclei are very far apart ($R_n, R_p \ll R_{_{\rm WS}}$) and since the nuclear interactions are short range
the screening is almost complete. Consequently
the effects of the nuclear clusters on the free neutron gas are negligible for bulk properties involving all free or valence states. This is illustrated on figure \ref{fig.eos}
 for the total energy density $\varepsilon$ of the free neutrons given by
 \begin{equation}\label{eos}
\varepsilon\{ n_f \}=\int_0^\mu {\cal E} {\cal N}\{ {\cal E} \} {\rm d}{\cal E} \, , \hskip 0.5cm n_f = \int_0^\mu {\cal N} \{ {\cal E} \} {\rm d}{\cal E}
 \end{equation}
where ${\cal N}\{ {\cal E} \}$ is the density of single particle states defined by an integral over the constant energy surface ${\cal E}_{\kk} = {\cal E}$
 \begin{equation}\label{dos}
{\cal N}\{ {\cal E} \} =\frac{2}{(2\pi)^3} \sum_\alpha \int \frac{ {\rm d} S^{(\alpha)}_{\cal E} }{| \boldsymbol{\nabla}_{\kk} {\cal E}_{\kk\alpha} | } \, .
 \end{equation}
 It can be seen on figure \ref{fig.eos} that the total energy density of the free neutrons does not depend neither on the chosen Skyrme interaction
 nor on the presence of the clusters and is actually very close to that of the ideal Fermi gas.
 We have computed the Fermi surface integral (\ref{dos}) for each energy on a grid of 20000 points in the range  $0<{\cal E}<0.3$ MeV.
 Each of the Fermi surface integrations were performed with 9920 cubes in the irreducible domain, \textit{i.e.} 476160 cubes in the first B-Z.
 Increasing the number of cubes did not lead to any noticeable change on the figure.

\begin{figure}
\centering
\epsfig{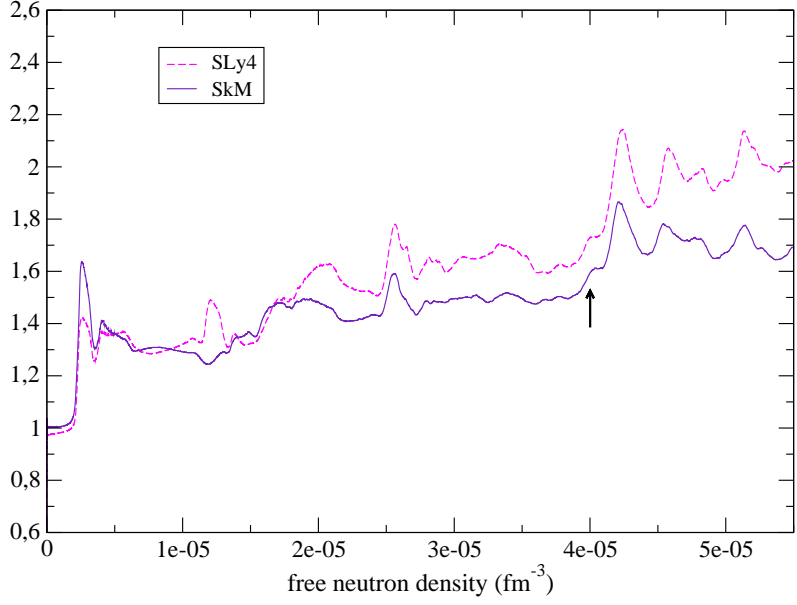}
\caption{Effective neutron mass $m_\star/m_n$ as a function of the free neutron density $n_f$ (in fm$^{-3}$).
 The arrow indicates the actual free neutron density in the W-S cell considered by Negele \& Vautherin.}
\label{fig.effmass}
\end{figure}

On the other hand, transport properties of the neutron gas depend only on the states lying on the
Fermi surface whose shape is determined by the interactions between the clusters and the neutron gas (unlike its volume
 which is simply given by ${\cal V}_{\rm F} = (2\pi)^3 n_n)$ as in the ideal Fermi gas).
As in the preceeding work \cite{Chamel05}, the density $n_f$ of free neutrons is defined by counting all single particle states
whose energy is larger than the maximum value of the mean field potential $U_n\{ \rr \}$.
As can be seen on figure \ref{fig.effmass} the effective neutron mass $m_\star$ deviates significantly from the ``bare'' neutron mass $m_n$
(as in the case of electrons in copper previously discussed in section \ref{Copper}),
and exhibits large variations due to band effects, except at very low dripped neutron densities $n_f$ for which $m_\star \sim m_n$.

\begin{figure}
\centering
\epsfig{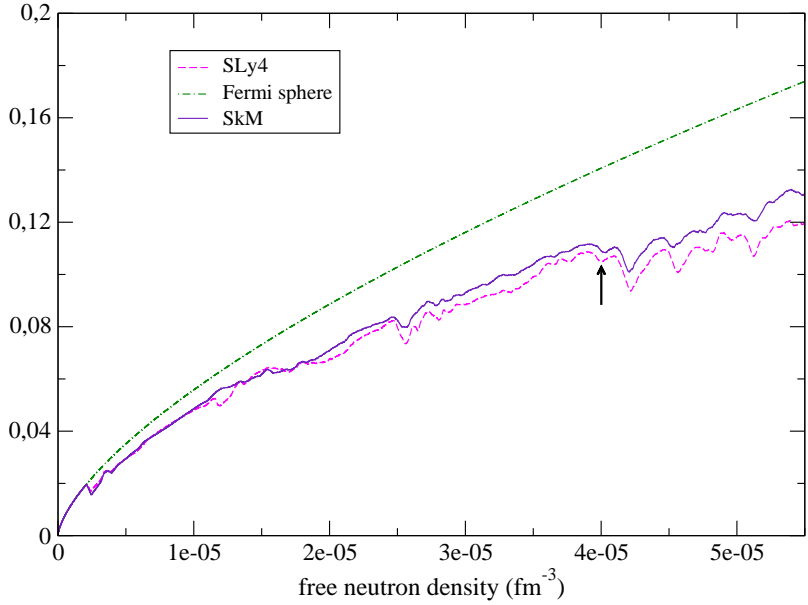}
\caption{Fermi surface area (in fm$^{-2}$) as a function of the free neutron density $n_f$ (in fm$^{-3}$) compared to the area of the corresponding
Fermi sphere. The arrow indicates the actual free neutron density in the W-S cell considered by Negele \& Vautherin.}
\label{fig.area}
\end{figure}

This can be understood from the consideration that the effects of the nuclear clusters are very small whenever the Fermi wave length
is much larger than the lattice spacing, or in other words when the Fermi volume lies well inside the first B-Z.
This means that the Fermi surface is nearly spherical for densities much lower than the density for which the Fermi sphere
touches the B-Z zone boundaries. Such a situation occurs for alkali metals.
For a body centered cubic lattice, this density threshold is given by
$n_f=\sqrt{2} \pi/3 {\cal V}_{\rm cell}$ (at this density, the radius of the Fermi sphere equals the
distance from the center of the first B-Z to the center of a face),
which yields $n_f\simeq 2.2 \times 10^{-6}$ fm$^{-3}$ for $R_{_{\rm WS}}=54$ fm.
Distorsions of the Fermi sphere are expected to occur for slightly lower densities due to the formation of necks as in copper (see figure \ref{fig.Cu}).
As shown on figure \ref{fig.area}, above this density the Fermi surface area is reduced compared to the area of the corresponding sphere
because by breaking the full translational symmetry (thereby raising some degeneracies), the presence of the nuclear lattice
leads to the opening of holes on the Fermi surface (via Bragg scattering). In the limit for which the
translational symmetry is completely lost (as in the case of isolated nuclei or in the W-S approximation),
the Fermi surface disappears and its area vanishes which obviously implies that ${\cal K}=0$.
The mobility coefficient can also vanish whenever the energy spectrum exhibits a gap at the Fermi level. This situation is realised in ordinary insulators
and in semiconductors for which the electric conductivity $\sigma=e^2 \tau {\cal K}=0$ since ${\cal K}=0$ (at zero temperature).
However this is still an open issue in the case of neutron star crust.

\section{Conclusion}

The band theory has been applied to investigate the transport properties of the neutron superfluid in the outermost layers of
neutron star crust, at densities around $4\times 10^{11}$ g.cm$^{-3}$. We have described the free neutrons with an effective
Schr\"odinger equation, involving a mean local mass $m_n^\oplus\{ \rr \}$ and a
mean potential $U_n\{ \rr \}$ deduced from the results of Negele\&Vautherin \cite{NV73} in
the extended Thomas-Fermi approximation with Skyrme nucleon-nucleon interactions. Unlike the calculations carried out by these
authors, we have solved the equations beyond the Wigner-Seitz approximation by applying Bloch boundary conditions.

It has been found that the effects of the nuclear clusters are very small when the neutron gas is very dilute, meaning that
the Fermi wavelength is much larger than the lattice spacing. A similar case is observed in alkali metals.
However at higher densities, the effective neutron mass $m_\star$ (which is analog to the electron optical effective mass in solid state
physics) is significantly increased compared to the ``bare'' mass due to Bragg scattering
(as it is also observed for the case of electrons in ordinary metals) and fluctuates
with the density as a result of band effects. The results also show significant differences with the choice of the parametrisation
of the Skyrme interaction. A more accurate evaluation of the effective neutron mass in the crust therefore requires
a fully self-consistent treatment in which both the equilibrium structure of the crust and the
transport properties are computed within the same nuclear model. However the main conclusions drawn in this paper are not expected
to be altered.

The present calculations performed in the outermost layers of the inner crust combined with those
 carried out recently in the bottom layers \cite{CCHI, Chamel05}, suggest that the effective neutron mass $m_\star/m_n$ could
 take very large values in the middle layers of the crust where the
 size of the nuclear clusters is of the same order as the lattice spacing. This raises the question whether in some layers
 the effective neutron mass actually diverges (in the sense that the mobility vanishes ${\cal K}=0$)
 due to the existence of band gaps in the energy spectrum. This question
 is one of the main issues in the study of any periodic materials such as for instance photonic or phononic crystals
  which have triggered considerable interests, both experimentally and theoretically.
In our calculation, no such gaps were found for the outermost layers of the inner crust however as
shown by Economou \textit{et al.} \cite{Economou89} in the context of photonic crystals, this does not exclude the
possibility of gaps at higher densities (higher energies), in deeper layers.
The existence of band gaps would have important implications for the dynamics of the star and for its cooling properties.

 Very large values of the effective mass
 have important consequences for the understanding of pulsar glitches. These
sudden increases in the rotational frequency of some pulsars are interpreted
as discontinuous transfers of angular momentum between the neutron superfluid and the solid crust. The present
results suggest that in the middle layers the neutron superfluid would be strongly coupled to the crust
so that such layers would contribute very little to the transfer of angular momentum. In the extreme case of band gaps,
the superfluid would be locked to the solid parts so that relative currents could not develop (for the same reasons that
electric currents do not flow in ordinary insulators).
Large enough effective masses could also trigger a Kelvin-Helmholtz instability \cite{Andersson04}, which could provide
a possible explanation for the origin of pulsar glitches. Further observational as well as theoretical investigations are required
in order to sheld light on the nature of pulsar glitches and the link with the physics of ``neutronic'' crystals in neutron star crust.

\bigskip
{\bf Acknowledgements}
\medskip

The author acknowledges financial support from the Lavoisier program of the French Ministry of Foreign Affairs. The author
is also very grateful to professor Haensel for discussions.

\appendix

\section*{Appendix 1: effective electron mass in metals}

Effective electron optical mass for a few elements taken from the compilation of reference \cite{Huttner96}.

\bigskip

\begin{tabular}{|p{1.5cm} | p{1.5cm} |  }\hline
metal & $m_\star/m_e$   \\ \hline
Ag & 0.99-1.04 \\ \hline
Al & 1.52  \\ \hline
Au & 1.1 \\ \hline
Be & 0.46 \\ \hline
Ni & 1.7  \\ \hline
Ca & 1.8 \\ \hline
Cr & 1.72 \\ \hline
Fe & 1.9 \\ \hline
Cu & 1.45 \\ \hline
Ir & 1.41 \\ \hline
K & 1.08 \\ \hline
Li & 1.45 \\ \hline
Pb & 2.12-2.71 \\ \hline
Mo & 1.43-2 \\ \hline
Na & 0.98-1.25 \\ \hline
Nb & 1.86 \\ \hline
Pd & 1.66 \\ \hline
Rh & 1.87 \\ \hline
Ta & 1.65 \\ \hline
Ti & 1.39 \\ \hline
V & 1.63 \\ \hline
W & 1.31 \\ \hline
\end{tabular}

\section*{Appendix 2: LAPW matrix elements}

We have extended the LAPW method as discussed in the solid state context by Koelling and Arbman \cite{Koelling75}
in order to allow for a non uniform mass in the kinetic term. The end result is simply the replacement
of the electron mass $m_e$ by the Skyrme neutron mass $m_n^\oplus\{ R_s \}$ in the expression of the matrix elements.
In reference \cite{Koelling75}, the authors used atomic units in which the kinetic factor $\hbar^2/2 m_e$ is equal to unity and therefore the electron mass
does not appear explicity. We report in this appendix the full expressions for the matrix elements.

The radial solution being normalised as
\begin{equation}
\int_0^{R_s} {\rm d}r\, r^2 u_\ell\{ r\}^2 = 1\, ,
\end{equation}
the Hamiltonian and the overlap matrix elements are given by
\begin{equation}
\begin{split}
\langle \phi_\alpha \vert \hat H \phi_\beta \rangle &=
\frac{\hbar^2 {\bf q_\alpha}\cdot {\bf q_\beta}}{2 m_n^{_\oplus}\{ R_s \}} \left(\delta_{\alpha\beta}
-\frac{4\pi R_s^2}{{\cal V}_{\rm cell}} \frac{j_1\{ |{\bf q_\alpha} - {\bf q_\beta}| \}}{ |{\bf q_\alpha} - {\bf q_\beta}| }\right) \\
&+\left( \frac{\hbar^2 R_s^2}{2 m_n^{_\oplus}\{ R_s \}}\right)^2\frac{4 \pi} {{\cal V}_{\rm cell}}   \sum_{\ell=0}^{+\infty} (2\ell+1) P_\ell\{\cos \theta\}
\left[ {\cal E}_\ell (a_\ell\{ q_\alpha \} a_\ell\{ q_\beta \}  + N_\ell b_\ell\{ q_\alpha \} b_\ell\{ q_\beta \} )
+ \gamma^\ell_{\alpha \beta} \right]
\end{split}
\end{equation}
\begin{equation}
\begin{split}
\langle \phi_\alpha \vert \phi_\beta \rangle&=
\delta_{\alpha\beta}-\frac{4\pi R_s^2}{{\cal V}_{\rm cell}} \frac{j_1\{ |{\bf q_\alpha} - {\bf q_\beta}| \}}{ |{\bf q_\alpha} - {\bf q_\beta}| } \\
&+\frac{4 \pi R_s^2} {{\cal V}_{\rm cell}} \sum_{\ell=0}^{+\infty}
\left(a_\ell\{ q_\alpha \} a_\ell\{ q_\beta \}  + N_\ell b_\ell\{ q_\alpha \} b_\ell\{ q_\beta \} \right)
\end{split}
\end{equation}
where $j_\ell$ is the regular spherical Bessel function of order $\ell$, $P_\ell$ is the Legendre polynomial of order $\ell$,
$\theta$ is the angle between the vectors $\pmb{q}_\alpha\equiv \kk+\pmb{K}_\alpha$ and $\pmb{q}_\beta=\kk+\pmb{K}_\beta$
(where $\pmb{K}_\alpha$ and $\pmb{K}_\beta$ are reciprocal lattice vectors),
\begin{equation}
N_\ell = \int_0^{R_s} {\rm d}r\, r^2 \dot{u}_\ell\{ r\}^2
\end{equation}
\begin{equation}
a_\ell\{ q \} = \dot{u}_\ell\{ R_s \} j_\ell^\prime\{q R_s \} - \dot{u}^\prime_\ell\{ R_s \} j_\ell\{ q R_s \}
\end{equation}
\begin{equation}
b_\ell\{ q \} = u^\prime_\ell\{ R_s \} j_\ell\{q R_s \} - u_\ell\{ R_s \} j^\prime_\ell\{ q R_s \}
\end{equation}
\begin{multline}
\gamma^\ell_{\alpha\beta} = \dot{u}_\ell \{ R_s \} u^\prime_\ell\{ R_s\} \biggl( j_\ell^\prime\{ q_\alpha R_s \} j_\ell\{ q_\beta R_s \}
+ j_\ell^\prime\{ q_\beta R_s \} j_\ell\{ q_\alpha R_s \}\biggr) \\- \dot{u}^\prime_\ell\{ R_s \} u^\prime_\ell\{ R_s \} j_\ell\{ q_\alpha R_s \}
 j_\ell\{ q_\beta R_s \} -\dot{u}_\ell\{ R_s \} u_\ell\{ R_s \} j^\prime_\ell\{ q_\alpha R_s \}
 j^\prime_\ell\{ q_\beta R_s \} \, .
\end{multline}

The prime means derivative with respect to the radial coordinate $r$.

\section*{Appendix 3: Skyrme parametrizations}

The values of the parameters of the Skyrme forces used in this work are given in the following table (the unit of energy is MeV and the
unit of length is fm)

\bigskip

\begin{tabular}{|p{1.5cm} | p{1.5cm} | p{1.5cm} |  }\hline
& SkM & SLy4  \\ \hline
$t_0$ & -2645 & -2488.91 \\ \hline
$t_1$ & 385 & 486.82 \\ \hline
$t_2$ & -120 &  -546.39 \\ \hline
$t_3$ & 15595 & 13777 \\ \hline
$x_0$ & 0.09 &  0.834 \\ \hline
$x_1$ & 0  & -0.3438 \\ \hline
$x_2$ & 0 & -1 \\ \hline
$x_3$ & 0 &  1.354 \\ \hline
$\gamma$ & 1/6 & 1/6 \\ \hline
\end{tabular}


\begin{thebibliography}{99}

\bibitem{Hae01} P. Haensel,
``Neutron Star Crusts'' in LNP Vol. 578: Physics of Neutron Star Interiors,
ed. D. Blaschke, N. K. Glendenning and A. Sedrakian,
Springer (2001), 127-174.

\bibitem{Baym69} G. Baym, C. Pethick, D. Pines, M. Ruderman,
``Spin up in neutron stars: the future of the Vela pulsar,
{\it Nature} {\bf 224} (1969) 872.

 \bibitem{CCHII} B. Carter, N. Chamel, P. Haensel,
``Entrainment coefficient and effective mass for conduction neutrons in neutron star crust: Macroscopic treatment'',
arXiv preprint [astro-ph/0408083].

\bibitem{ChamelCarter05} N. Chamel, B. Carter,
``Effect of entrainment on stress and pulsar glitches in stratified neutron star crust'',
arXiv preprint [astro-ph/0503044].

\bibitem{CCHI} B. Carter, N. Chamel, P. Haensel,
``Entrainment coefficient and effective mass for conduction neutrons
in neutron star crust: simple microscopic models'',
{\it Nucl. Phys.} {\bf A748} (2005) 675-697.

\bibitem{Chamel05} N. Chamel, ``Band structure effects for dripped neutrons
in neutron star crust'',
{\it Nucl. Phys A} {\bf 747} (2005) 109-128.

\bibitem{NV73} J. W. Negele, D. Vautherin,
``Neutron star matter at sub nuclear densities'',
{\it Nucl. Phys. A} {\bf 207} (1973), 298-320.



\bibitem{CCHIII} B. Carter, N. Chamel, P. Haensel,
``Effect of BCS pairing on entrainment in neutron superfluid current
in neutron star crust'',
{\it Nucl. Phys. A} {\bf 759} (2005) 441-164.

\bibitem{Baldo05} M. Baldo, E. E. Saperstein, S. V. Tolokonnikov,
``Superfluidity in nuclear and neutron matter'',
{\it Nucl. Phys A} {\bf 749} (2005) 42c-52c.

\bibitem{Cohen58} M. H. Cohen,
``Optical constants, heat capacity, and the Fermi surface'',
{\it Phil. Mag.} {\bf 49} (1958) 762.

\bibitem{Huttner96} B. H\"uttner,
``A new method for the determination of the optical mass of electrons in metals'',
{\it J. Phys.:Condens. Matter } {\bf 8} (1996) 11041-11052.

\bibitem{Kittel} C. Kittel,
``Introduction to solid state physics'',
John Wiley \& sons, 7th edition (1996).

\bibitem{Zeilinger86} A. Zeilinger, C. G. Shull, M. A. Horne, K. Finkelstein,
``Effective mass of neutrons diffracting in crystals'',
{\it Phys. Rev. Lett. }{\bf 57} (1986) 3089-3092.

\bibitem{Brack85} M. Brack, C. Guet, H. B. Hakansson,
``Selfconsistent semiclassical description of average nuclear properties-a link between microscopic and macroscopic models'',
{\it Phys. Rep. } {\bf 123} (1985) 275-364.

\bibitem{SkM} H. Krivine, J. Treiner, O. Bohigas,
``Derivation of a fluid-dynamical lagrangian and electric giant resonances'',
{\it Nucl. Phys. A} {\bf 336} (1980) 155-184.

\bibitem{Siemens71} P. J. Siemens, V. R. Pandharipande,
``Neutron matter computations in Bruckner and variational theories'',
{\it Nucl. Phys. A} {\bf 173} (1971) 561-570.

\bibitem{Bonche81} P. Bonche, D. Vautherin,
``A mean field calculation of the equation of state of supernova matter'',
{\it Nucl. Phys. A} {\bf 372} (1981) 496-526.

\bibitem{SLy4} E. Chabanat, P. Bonche, P. Haensel, J. Meyer, R. Schaeffer,
``New Skyrme effective forces for supernovae and neutron rich nuclei'',
{\it Physica Scripta} {\bf T56} (1995) 231.

\bibitem{SLy4b} E. Chabanat, P. Bonche, P. Haensel, J. Meyer, R. Schaeffer,
``A Skyrme parametrization from subnuclear to neutron star densities'',
{\it Nucl. Phys. A} {\bf 627} (1997) 710-746.

\bibitem{SLy4c} E. Chabanat, P. Bonche, P. Haensel, J. Meyer, R. Schaeffer,
``A Skyrme parametrization from subnuclear to neutron star densities Part II. Nuclei far from stabilities'',
{\it Nucl. Phys. A} {\bf 635} (1998) 231-256 ; Erratum, {\it Nucl. Phys. A} {\bf 643} (1998) 441.

\bibitem{Douchin00} F. Douchin, P. Haensel, J. Meyer,
``Nuclear surface and curvature properties for SLy Skyrme forces
and nuclei in the inner neutron star crust'',
{\it Nucl. Phys. A} {\bf 665} (2000) 419-446.

\bibitem{DouchinHaensel00} F. Douchin, P. Haensel,
``Inner edge of neutron-star crust with SLy effective nucleon-nucleon interactions'',
{\it Phys. Lett. B} {\bf 485} (2000) 107-114.

\bibitem{Magier2003} P. Magierski, A. Bulgac, P.-H. Heenen,
``Exotic nuclear phases in the inner crust of neutron stars in the light
of the Skyrme-Hartree-Fock theory'',
{\it Nucl. Phys. A} {\bf 719} (2003) 217-220.

\bibitem{Sandulescu04} N. Sandulescu, N. van Giai, R. J. Liotta,
``Superfluid properties of the inner crust of neutron stars'',
{\it Phys. Rev. C} {\bf 69} (2004) 045802.

\bibitem{Sandulescu04b} N. Sandulescu,
``Nuclear superfluidity and specific heat in the inner crust of neutron stars'',
{\it Phys. Rev. C} {\bf 70} (2004)025801.

\bibitem{Khan05} E. Khan, N. Sandulescu, N. van Giai,
``Collective excitations in the inner crust of neutron stars: Supergiant resonances'',
{\it Phys. Rev. C} {\bf 71} (2005) 042801.

\bibitem{Singh94} D. J. Singh,
``Planewaves, pseudopotentials and the LAPW method'',
Kluwer Academic Publisher, Boston, Dordrecht, London, (1994).

\bibitem{Slater37} J. C. Slater,
``Wave Functions in a Periodic Potential'',
{\it Phys. Rev. } {\bf 51} (1937) 846-851.

 \bibitem{Koster57} G. F. Koster, ``Space Groups and Their Representations'',
{Solid State Physics}, {\bf Vol 5}, F. Seitz \& D. Turnbull editors,
Academic Press Inc., Publishers New York (1957).

\bibitem{GilatRaubenheimer66} G. Gilat \& L.J. Raubenheimer
``Accurate Numerical Method for Calculating Frequency distribution Functions in Solids'',
{\it Phys. Rev.}{\bf V144, N2} (1966), 390-395 ; erratum: G. Gilat \& L.J. Raubenheimer
``Accurate Numerical Method for Calculating Frequency distribution Functions in Solids'',
{\it Phys. Rev.}{\bf V147, N2} (1966), 670.

\bibitem{Koelling75} D. D. Koelling \& G. O. Arbman,
``Use of energy derivative of the radial solution in an augmented plane wave method:
application to copper'',
{\it J. Phys. F} {\bf 5} (1975) 2041-2054.



\bibitem{Feynman39} R. Feynman,
``Forces in molecules'',
{\it Phys. Rev.} {\bf V56-I4} (1939) 340-343.

 \bibitem{Shockley37} W. Shockley, ``The Empty Lattice Test of the Cellular Method in Solids'',
 {\it Phys. Rev. }{\bf 52} (1937), 866-872.

 \bibitem{Choy00} T.-S. Choy, J. Naset, J. Chen, S. Hershfield, and C. Stanton,
 ``A database of fermi surface in virtual reality modeling language (vrml)'',
 {\it Bull. Am. Phys. Soc. }{\bf 45} (2000) L36 42.

\bibitem{Burdick63} G. A. Burdick,
``Energy band structure of Copper'',
{\it Phys. Rev.} {\bf 129} (1963) 138-150.

\bibitem{Seagall61} B. Segall,
``Fermi surface and energy bands of Copper'',
{\it Phys. Rev.} {\bf 125} (1962) 109-122.



\bibitem{PhillipsKleinman59} J. C. Phillips, L. Kleinman,
``New Method for Calculating Wave Functions in Crystals and Molecules'',
{\it Phys. Rev.} {\bf 59} (1959) 287-294.

\bibitem{Andersson04} N. Andersson, G. L. Comer, R. Prix,
``The superfluid two-stream instability'',
{\it Month. Not. Roy. Soc.} {\bf 354} (2004) 101-110.

\bibitem{Economou89} E. N. Economou, A. Zdetsis,
``Classical wave propagation in periodic structures'',
{\it Phys. Rev. B} {\bf 40} (1989) 1334-1337.



\end{thebibliography}
\end{document}